\documentclass{article}
\usepackage{epsfig}

\newcommand{\totcs}{$\sigma_{\rm tot}$ }
\newcommand{\elcs}{$\sigma_{\rm el}$ }
\newcommand{\diffcs}{$\sigma_{\rm diff}$ }
\newcommand{\ie}{{\it i.e. }}

\title{Diffraction at Tevatron and LHC\\ in the Miettinen-Pumplin model
\thanks{Presented at the XLV Cracow School of Theoretical Physics, Zakopane, Poland, June 3-12, 2005}
}

\author{S.~Sapeta \\ \\
\textit{M. Smoluchowski Institute of Physics, Jagellonian University}  \\
\textit{Reymonta 4, 30-059 Krak\'ow, Poland}
}

\date{}

\begin{document}
\maketitle
\begin{abstract}
The process of soft diffractive dissociation in hadronic collisions is discussed in the framework of the Miettinen-Pumplin model. A good description of the data in the ISR-Tevatron energy range is found. Predictions for the total, elastic and single diffractive cross sections for the LHC are also presented. The total cross section  is expected to be $15 \%$  smaller than that given by Donnachie and Landshoff in the model with soft pomeron. The  diffractive cross section remains constant in the Tevatron-LHC energy range. 
\end{abstract}

\section{Introduction}
We are interested in a diffractive process $pp \to pX $,
in which one of the colliding protons remains intact. The other dissociates into a system of particles well separated in rapidity from the intact proton.
The diffraction is called soft if there is no hard scale involved, \ie all transverse momenta of the final state particles are much smaller than the proton mass. The self-consistent description of this kind of processes is an important problem. The Regge theory is traditionally used to determine the cross sections. However, in the case of soft diffraction, the Regge approach based on the triple pomeron picture,  fails to describe the diffractive cross section for center of mass energies higher than about 20 {\rm GeV} \cite{Ingelman:1984ns, Kaidalov:jz}. 
This signals violation of unitarity in the pomeron approach which occurs for much lower energies than in the case of fully inclusive cross section. 
The way out of this problem was proposed some time ago by Goulianos who introduced, somewhat {\it ad~hoc}, the renormalization of the pomeron flux~\cite{Goulianos:wy}. 

However, diffractive dissociation may be also analyzed in the framework different than the Regge model. In this paper we present another approach proposed by Good and Walker \cite{Good:1960ba} which is based, from the very beginning, on the requirement of unitarity of the scattering matrix.
The results presented here are obtained using the Miettinen-Pumplin \cite{Miettinen:1978jb} realization of the~Good-Walker picture.

\section{The Miettinen-Pumplin model}

In the Good-Walker picture of soft diffraction the state of the incident hadron which subsequently dissociates is expanded into a superposition of eigenstates of the scattering operator ${\rm Im} T$
\begin{equation}
\mid B\rangle\,=\,\sum_{k}C_k\mid\psi_k \rangle, \label{expansion}
\end{equation}
\begin{equation}
{\rm Im} T \mid\psi_k \rangle\,=\,t_k\mid\psi_k \rangle\,,
\end {equation}
where from unitarity: $0\le t_k \le 1$.
In general case  different eigenstates are absorbed by the target with different
intensity, hence the outgoing state is no longer $\mid B \rangle$ and, by this mechanism, the inelastic production of particles takes place. The inelastic diffractive cross section is proportional to  the dispersion of the absorption coefficients $t_k$.

The Miettinen-Pumplin model is based on this simple picture of Good and Walker introducing new important element. The basic assumption is that the eigenstates of diffraction are wee parton states 
\begin{equation}
\mid \psi_k \rangle\equiv \mid \vec{b}_1,...,\vec{b}_N,y_1,...,y_N \rangle, \label{as}
\end {equation}
where $N$ is the number of partons, and $(y_i,\vec{b}_i)$ are rapidity and impact parameter (relative to the center of the projectile) of parton $i$,
respectively. Therefore, Eq.~(\ref{expansion}) takes the form
\begin{equation}
\mid B\rangle\,=\,\sum_{N=0}^{\infty}\int \prod_{i=1}^{N} d^2\vec{b}_i\,dy_i  \,C_N(\vec{b}_1,...,\vec{b}_N,y_1,...,y_N) \mid \vec{b}_1,...,\vec{b}_N,
\ y_1,...,y_N \rangle.
\end{equation}
The probability $\mid C_N \mid^2$ associated with $N$ partons, which are assumed to be independent, is given by Poisson distribution with mean number $G^2$
\begin{equation}
\mid C_N(\vec{b_1},...,\vec{b_N}, y_1,...,y_N)\mid^2 = e^{-G^2} \frac{G^{2N}}{N!}\prod_{i=1}^{N} \mid C(\vec{b_i}, y_i)\mid^2,
\end{equation} 
where $\mid C(\vec{b_i}, y_i)\mid^2$ is the single wee parton distribution probability. Similarly the interaction probability $t_k$ of the state with $N$ partons can be expressed in terms of the single wee parton interaction probability $\tau(\vec{b}_i,y_i)$
\begin{equation}
t_N(\vec{b_1},...,\vec{b_N}, y_1,...,y_N)=1-\prod^N_{i=1}(1-\tau(\vec{b}_i,y_i)).
\end{equation}
To describe distribution and interactions of single wee partons Miettinen and Pumplin took  
\begin{equation}
\mid C(b_i,y_i)\mid^2 = \frac{1}{2\pi\beta\lambda}
\exp\left(-\frac{\mid y_i\mid}{\lambda}-\frac{b_i^2}{\beta}\right),
\end{equation}
\begin{equation}
\tau(b_i,y_i)= A\exp\left(-\frac{\mid y_i\mid}{\alpha}-\frac{b_i^2}{\gamma}\right).
\end{equation}
With some further assumptions the number of parameters of the model may be reduced so that it depends only on $\beta[{\rm fm}^2]$ and $G^2$. Namely, $A=1$, its maximal possible value while $\alpha/\lambda=2.0$ and $\gamma/\beta=2.0$  (see \cite{Miettinen:1978jb}). Moreover, it turns out that $\alpha$ and $\lambda$ enter only as their ratio. Finally, we arrive at the following formulae for the differential total, elastic and single diffractive cross sections
\begin{eqnarray}
\frac{d\sigma_{\rm tot}}{d^2b}\,=\,
2\left(1-\exp\left(-G^2\frac{4}{9}\,
\,e^{-b^2/(3\beta)}\right)\right),
\label{eq:tot}
\end{eqnarray}
\begin{eqnarray}
\frac{d\sigma_{\rm el}}{d^2b}
\,=\,
\left(1-\exp\left(-G^2\,\frac{4}{9}\,e^{-b^2/(3\beta)}\right)\right)^2,
\label{eq:el}
\end{eqnarray}
\begin{eqnarray}
\frac{d\sigma_{\rm diff}}{d^2b}
\,=\,
\exp\left(-2\,G^2\,\frac{4}{9}\, e^{-b^2/(3\beta)}\right) \left(\exp\left(G^2\,\frac{1}{4}\,e^{-b^2/(2\beta)}\right)-1\right)\,.
 \label{eq:diff}
\end{eqnarray}
The two remaining parameters, $\beta$ and $G^2$, can be determined for a given center of mass energy  $\sqrt{s}$ from experimental data for \totcs and \elcs using Eqs.~(\ref{eq:tot}) and (\ref{eq:el}). The diffractive cross section can be then {\it predicted} from Eq.~(\ref{eq:diff}).

Miettinen and Pumplin performed calculations for two colliding
protons at the ISR center of mass energy $\sqrt{s}= {\rm 53\ GeV}$. They obtained the value for \diffcs which was in good agreement with the data. We present this result in Fig.~\ref{fig:tevatron-diff} and refer to it as M\&P.

\section{Diffraction at Tevatron}

\begin{figure}[t]
\begin{center}
\epsfig{file=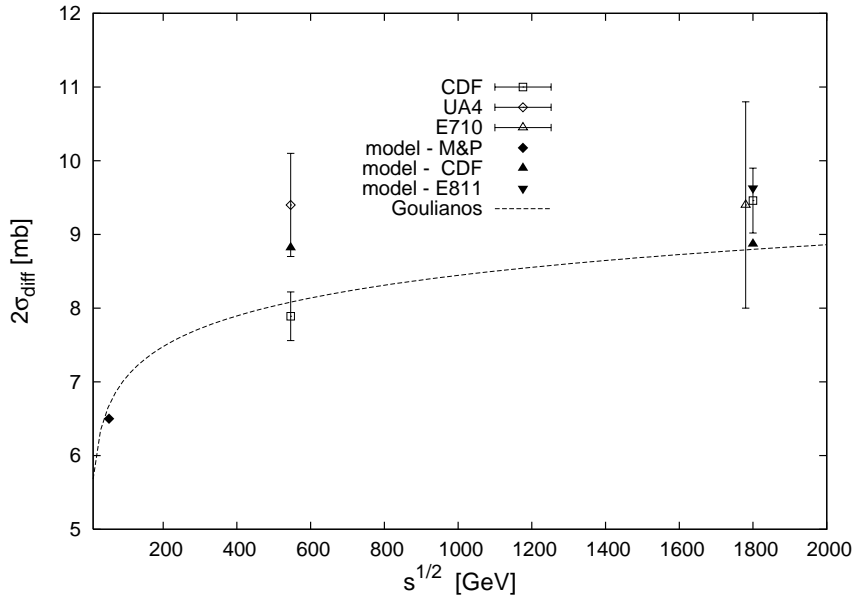, width=8cm, angle=-90}
\end{center}
\vskip -0.2cm
\caption{Diffractive cross section for high energies. The open points represent available experimental data for diffractive dissociation. The black points are the predictions of the Miettinen-Pumplin model. The dashed line refers to the Goulianos model.}
\label{fig:tevatron-diff}
\end{figure}

We have applied the model described in the previous section for the center of mass energies 546 GeV and 1800  GeV \cite{Sapeta:2004av}. The results together with experimental data and the Goulianos model predictions are shown in Fig.~\ref{fig:tevatron-diff}.
We used CDF~\cite{Abe:1993xy} data for the total and elastic cross sections as an input at the energy 546 GeV.
The two predictions of the Miettinen-Pumplin model for $\sqrt{s}= {\rm 1800 \ GeV}$ are a consequence of two different results for \totcs and \elcs measured by CDF~\cite{Abe:1993xy} and E811~\cite{Avila:1998ej}. 

We see that the Miettinen-Pumplin model remains valid in the ISR-Tevatron energy range, \ie for three orders of magnitude in the center of mass energy $\sqrt{s}$. It gives values of diffractive cross section which are in reasonable agreement with the data.

It is also possible to determine within the model the elastic and diffractive slopes by applying Fourier transform to Eqs. (\ref{eq:el}) and (\ref{eq:diff}). We have checked that both slopes are consistent with existing experimental data (see~\cite{Sapeta:2004av, Sapeta:2005ba}) which undoubtly makes the Miettinen-Pumplin model more trustworthy.

\section{Predictions for LHC}

\begin{figure}[t]
\begin{center}
\epsfig{file=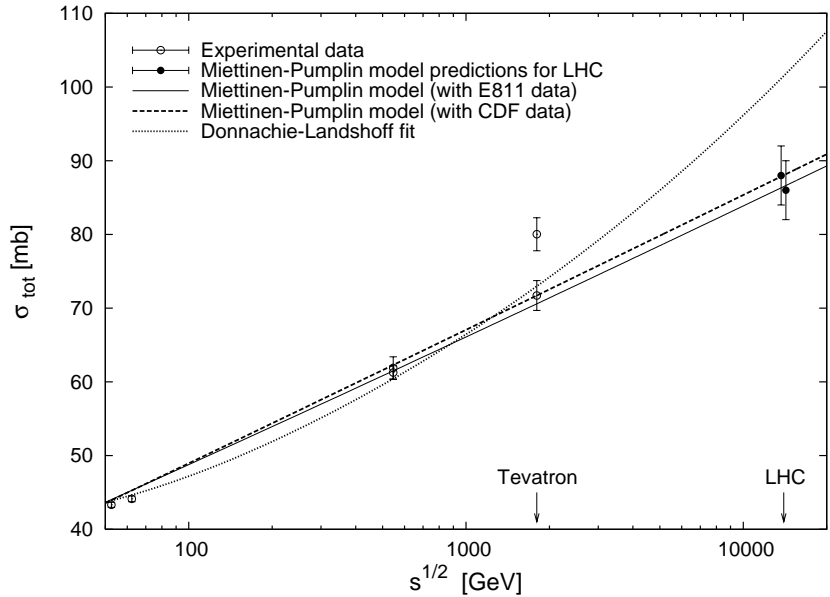, width=8cm, angle=-90}
\end{center}
\vskip -0.2cm
\caption{Total cross section from the Miettinen-Pumplin model together with the Donnachie-Landshoff prediction. Data points at the LHC energy are predictions  from
Table~1. Experimental data are from \cite{Avila:1998ej, Abe:1993xy, Amos:1985wx, Bozzo:1984rk}.}
\label{fig:lhc-tot}
\end{figure}

Encouraged by the success of the Miettinen-Pumplin model for the highest currently available energy of Tevatron, we have proposed a method of determining the total, elastic and diffractive cross sections at the LHC \cite{Sapeta:2005ba}.

In order to obtain these predictions we have extrapolated the two parameters of the model $\beta$  and $G^2$ to the LHC energy $\sqrt{s}=14~{\rm TeV}$. For this purpose we plotted the obtained values of $\beta$ and $G^2$ as a functions of energy and found that up to the Tevatron value of $\sqrt{s}$ the dependence of both parameters is, to good approximation, linear in $\ln\sqrt{s}$.
Thus we extrapolated this dependence to the LHC energy by fitting straight lines to the existing data points. 
It is interesting to note that with the assumption of the linear dependence, the total cross section for high $\sqrt{s}$ behaves like  
\begin{equation}
\sigma_{\rm tot}\, \propto\, \ln(s)\,\ln(\ln s)\,, \label{froissart}
\end{equation}
which is smaller than $\ln^2 s$ and therefore does not violate the Froissart-Martin bound \cite{FMbound}.

\begin{figure}[ht]
\begin{center}
\epsfig{file=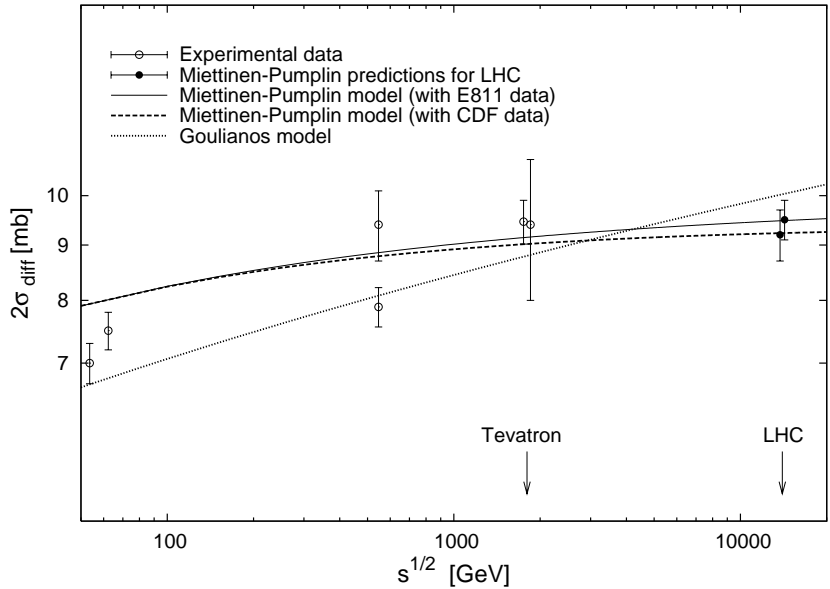, width=8cm, angle=-90}
\end{center}
\vskip -0.2cm
\caption{Diffractive cross section from the Miettinen-Pumplin model together with the prediction of Goulianos . Data points at the LHC energy are predictions from Table~1. Experimental data are from \cite{Amos:1992jw, Armitage:1982, Bernard:1986yh, Abe:1993wu}.}
\label{fig:lhc-diff}
\end{figure}

When fitting the energy dependence, we faced the problem pointed out already in the previous section, \ie  discrepancy between E811 and CDF results for \totcs and \elcs. Thus we decided to treat these two cases separately considering two scenarios. The results are presented in Table \ref{cs-results}.
The indicated errors come from uncertainties in the determination of parameters and were computed by using the total differential method. 
The dependence of the total and diffractive cross sections on the center of mass energy  is shown in Figs. \ref{fig:lhc-tot} and \ref{fig:lhc-diff}.

\begin{table}[ht]
\begin{center}
\begin{tabular}{|c|c|c|c|} \hline
 Scenarios     & \totcs~[mb]  & \elcs~[mb] & \diffcs~[mb] \\ \hline
with E811 data \cite{Avila:1998ej}  &    $86  \pm 4$    &   $21  \pm 1$   &    $9.5  \pm 0.4$  \\ \hline
with CDF data  \cite{Abe:1993xy}  &   $88  \pm 4$    &   $22  \pm 2$   &    $9.2  \pm 0.5$ \\ \hline
\end{tabular}
\end{center}
\vskip -0.2cm
\caption{Predictions of the Miettinen-Pumplin model for the total, elastic and diffractive cross sections
at the LHC energy $14~{\rm TeV}$, calculated in two scenarios.} 
\label{cs-results}
\end{table}

As we see the Miettinen-Pumplin model with assumed logarithmic dependence of its two parameters $\beta$ and $G^2$ on center of mass energy  predicts the total cross section for the LHC 15\% smaller than that determined by Donnachie and Landshoff \cite{Donnachie:1992ny}. This difference can be attributed to unitarity which is an inherent feature of this model. The value of the diffracitve cross section at the LHC is only slightly higher than that found at Tevatron and is close to the prediction of the Goulianos model. Despite this simillary, the two models give qualitatively different behaviour of the diffractive cross section. The model of Miettinen and Pumplin predicts \diffcs almost constant in the Tevatron-LHC energy range while in the Goulianos model the diffractive cross section grows with energy.

\section{Summary}

We have analyzed the soft diffractive dissociation in hadronic collisions at high energies in the framework of the Miettinen-Pumplin model. 

We have found correct description of the single diffractive cross section at center of mass energies ranging from ISR to Tevatron. We have also presented predictions for the LHC, finding the total inclusive cross section 15\% smaller then that determined by Donnachie and Landshoff. The diffractive cross section is predicted to be almost constant in the Tevatron-LHC energy range. 

\vspace{0.5cm}
The original work described here was partly done in collaboration with Krzysztof Golec-Biernat. This research has been supported by the grants of Polish Ministry of Science and Information Society Technologies: No.~1~P03B~02828 and No.~2~P03B~04324.

\end{document}